\documentclass[prd,aps,preprintnumbers,nofootinbib,twocolumn]{revtex4}
\usepackage{epsfig}
\usepackage{amsmath}
\usepackage{hyperref}

\begin{document}

\preprint{Cavendish-HEP-25/01, IFJPAN-IV-2025-6, P3H-25-016, TTK-25-09}

\renewcommand{\thefigure}{\arabic{figure}}

\title{Identified Hadron Production at Hadron Colliders\\in Next-to-Next-to-Leading-Order QCD}

\author{Micha\l{}  Czakon}
\affiliation{{\small Institut f\"ur Theoretische Teilchenphysik und Kosmologie, RWTH Aachen University, D-52056 Aachen, Germany}}
\author{Terry Generet}
\author{Alexander Mitov}
\affiliation{{\small Cavendish Laboratory, University of Cambridge, Cambridge CB3 0HE, UK}}
\author{Rene Poncelet}
\affiliation{{\small The Henryk Niewodnicza\'nski Institute of Nuclear Physics, ul. Radzikowskiego 152, 31-342 Krakow, Poland}}

\date{\today}

\begin{abstract}
In this work we calculate for the first time the next-to-next-to leading order (NNLO) QCD corrections to identified hadron production at hadron colliders. The inclusion of the NNLO correction has an important impact on all observables considered in this work. Higher order corrections reduce scale uncertainty and in almost all cases are moderate. Overall, good perturbative convergence is observed across kinematics and observables. The uncertainty due to missing higher orders is relatively small and, in many cases, smaller than the experimental uncertainty. The largest source of theoretical uncertainty at present is from the knowledge of the non-perturbative parton-to-hadron fragmentation functions (FF), which dwarfs the scale uncertainty in most kinematic ranges. The inclusion of NNLO corrections demonstrates the precision studies potential of this class of observables. To fully realize this potential, however, a new generation of improved fragmentation functions may be needed. The results of the present work will enable global fits of FF with NNLO precision.
\end{abstract}

\maketitle

\section{Introduction}

The theory of the strong interactions, Quantum Chromodynamics (QCD), is a pillar of collider physics. Almost all high-energy collider measurements rely on QCD, either as a direct input to measurements, or as part of the interpretation of Standard Model (SM) measurements and searches for physics beyond the SM.

Given QCD's importance, it is natural to ask how one would test and validate QCD itself. To this end, ideally, one would like to compare QCD predictions with measurements of observables that are as little sensitive as possible to the electroweak part of the SM. A prime example is the class of observables based on identified light hadrons like $\pi^\pm,\pi^0,p$, etc.

Observables based on identified hadrons are sensitive to perturbative corrections, which are computable in perturbation theory. Such observables also provide direct access to the non-perturbative aspect of QCD via the so-called parton-to-hadron fragmentation functions (FF). A detailed study of the production of single and multiple identified hadrons allows one to study the FF universality and scaling violations, verify QCD factorization theorems and place quantitative constraints on non-perturbative power-suppressed corrections in collider processes.

Existing hadron collider studies have been restricted to next-to-leading order (NLO) in QCD \cite{Aversa:1988vb,Jager:2002xm}. Very recently soft-gluon resummation for this class of observables was extended to full next-to-next-to leading logarithmic (NNLL) accuracy \cite{Hinderer:2018nkb}. Hadron-in-jet observables, also considered in this work, are known through NLO and NLL \cite{Arleo:2013tya,Kaufmann:2019ksh,Liu:2023fsq,Caletti:2024xaw}. 

Presently, FF cannot be derived in QCD from first principles and have to be extracted from data. A principal source for fitting FF is semi-inclusive annihilation (SIA) $e^+e^-$ data. Semi-inclusive DIS (SIDIS) is another important source for extracting FF. The excellent LHC data taking ability makes the use of hadron collider data for extracting FF an attractive possibility.

Many FF sets exist at present \cite{Hirai:2007cx,deFlorian:2007ekg,Albino:2008fy,deFlorian:2014xna,Bertone:2017tyb,Bertone:2018ecm,Moffat:2021dji,Khalek:2021gxf,Borsa:2021ran,Borsa:2022vvp,AbdulKhalek:2022laj,Gao:2024dbv}. Almost all are with NLO accuracy since, until recently, only SIA data could be fitted with NNLO accuracy. Very recently, NNLO calculation of SIDIS appeared \cite{Goyal:2023zdi,Bonino:2024qbh,Goyal:2024tmo,Bonino:2024wgg,Ahmed:2024owh,Goyal:2024emo} which now makes possible the extraction of FF with NNLO precision also from DIS processes. The only major process not yet available with NNLO precision is identified hadron production in hadron collisions. This capability gap is a major obstacle to performing truly global FF fits with NNLO precision.

In this work we initiate a research program which aims to conclusively address some of the above questions. Specifically, we use recent advances in perturbative QCD calculations to produce, for the very first time, QCD predictions for single and multiple identified hadrons at hadron colliders with full next-to-next-to leading order (NNLO) precision. We clarify questions like the level of improvement the NNLO QCD correction brings to these observables, if higher order corrections can remove any remaining discrepancy between QCD and data and, finally, assess the possibilities for future extraction of FF with NNLO precision from hadron collider data. 

While our immediate goal is to improve the description of single and multiple identified hadron production, we hope that ultimately, our results will help place improved constraints on non-perturbative, power-suppressed corrections to hadron collider observables.

\section{Calculating the NNLO correction}

The differential cross-section for producing an identified light hadron $h$ with momentum $p$ in hadron collisions, $pp\to h+X$, reads
\begin{equation}
d\sigma_{pp\to h}(p) = \sum_{i} \int dz~ d\hat\sigma_{pp\to i}\left(\frac{p}{z}\right) D_{i\to h}(z)\,,
\label{eq:xsection}
\end{equation}
where $d\hat\sigma_{pp\to i}$ is the $\overline{\rm MS}$-subtracted partonic cross-section for producing a parton $i$ with momentum $p_i=p/z$. The sum in eq.~(\ref{eq:xsection}) runs over all massless partons $i$ which can contribute at the factorization scale implicit in eq.~(\ref{eq:xsection}). Throughout this work we use $n_f=5$ massless flavors. 

$D_{i\to h}$ is a set of non-perturbative fragmentation functions which we take from the literature. Specifically, in this work we use the NNFF \cite{Bertone:2017tyb} and MAPFF \cite{AbdulKhalek:2022laj} sets for pion final states, and the NNFF \cite{Bertone:2018ecm} and JAM \cite{Moffat:2021dji} FF sets for charged hadron observables. The FFs depend on the factorization scale and partonic fraction $z$. All FF sets used in this work are available as {\tt LHAPDF} library \cite{Buckley:2014ana} grids, which incorporate DGLAP evolution. Throughout this work we use the parton distribution function (PDF) set {\tt NNPDF3.1} \cite{NNPDF:2017mvq}, which is implicitly included in the partonic coefficient function $d\hat\sigma$. The strong coupling constant $\alpha_S$ is taken at NNLL as provided by the PDFs {\tt LHAPDF} grids.

The calculation of the NNLO massless coefficient function $d\hat\sigma_{pp\to i}$ is purely numeric. It was recently used in ref.~\cite{Czakon:2024tjr} in the NNLO calculation of open $B$ production at hadron colliders. The handling of infrared singularities at NNLO follows the {\tt STRIPPER} approach of refs.~\cite{Czakon:2010td,Czakon:2011ve} as implemented in refs.~\cite{Czakon:2014oma,Czakon:2019tmo}. The $\overline{\rm MS}$ subtraction of the final state collinear singularities associated with the final state parton $p_i$ is  performed numerically as explained in ref.~\cite{Czakon:2021ohs}. The required two loop four-parton massless amplitudes are taken from ref.~\cite{Broggio:2014hoa}, which are in agreement with the earlier calculations of refs.~\cite{Bern:2002tk,Bern:2003ck,Glover:2003cm,Glover:2004si,DeFreitas:2004kmi}. The required one loop amplitudes are taken from the library {\tt OpenLoops 2} \cite{Buccioni:2019sur}. Tree-level amplitudes are computed with the help of the {\tt AvH} library \cite{Bury:2015dla}.

In our calculations we set the central values of the factorization, renormalization and fragmentation scales to a common value. This value is chosen individually for each observable. The scale choices we make are specified in section~\ref{sec:pheno}. Scale variation is performed in the 15 point approach, where each scale is varied independently by a factor of two, omitting extreme variations where two of the scales differ by more than a factor of two.

Another novel feature of this work is the calculation of two-hadron inclusive observables at NNLO. Specifically, we provide predictions for the invariant mass $m(h_1,h_2)$ of two identified hadrons $h_1$ and $h_2$. Such an observable is defined through a straightforward extension of eq.~(\ref{eq:xsection})
\begin{eqnarray}
d\sigma_{pp\to h_1 h_2}(p_1,p_2) &=& \sum_{i,j} \int dz_1\int dz_2~ d\hat\sigma_{pp\to ij}\left(\frac{p_1}{z_1},\frac{p_2}{z_2}\right) \nonumber\\
&& \times~ D_{i\to h_1}(z_1)D_{j\to h_2}(z_2)\,,
\label{eq:xsection-m}
\end{eqnarray}
with $p_i=p_1/z_1$ and $p_j=p_2/z_2$. Note that this formula is valid as long as the invariant mass of the two hadrons is sufficiently large to suppress contributions from a single parton fragmenting to two hadrons.

The collinear singularities associated with each one of the identified final state partons $i,j$ is performed in the $\overline{\rm MS}$ scheme just like for the single inclusive case discussed above. Our results exhibits good numerical convergence and cancellation of all infrared singularities within the high numeric precision achieved in our calculations.

\section{LHC Phenomenology}\label{sec:pheno}

In this work we consider three different observables: the transverse momentum $p_T(\pi^0)$ of a neutral pion, the invariant mass $m(\pi^0,\pi^0)$ of two neutral pions, and the ratio $z=p_T(h)\cos(\Delta R)/p_T({\rm jet})$ of a charged hadron inside a jet, where $\Delta R = \sqrt{\Delta\phi^2+\Delta\eta^2}$ and $\Delta\phi$ and $\Delta\eta$ are the differences in azimuthal angle and pseudorapidity, respectively, between the identified hadron and jet momenta. Such a range of observables is sufficient for demonstrating the general behavior of NNLO QCD corrections in hadron collider observables with identified hadrons. All calculations shown in this work, independently of their perturbative order, use PDFs with NNLO accuracy. The accuracy of the FF is also kept fixed at their highest available order, NLO or NNLO, depending on the FF set. For all observables involving neutral pions, the $\pi^0$ fragmentation functions are derived by averaging the $\pi^+$ and $\pi^-$ FFs of the corresponding FF set. This assumption is often made by FF fitting collaborations and has been found, e.g.~in ref.~\cite{Gao:2025bko}, to be valid to very good approximation.

We make the following central scale choices for the three observables considered in this work: $\mu^2 = \max\left(16\, {\rm GeV}^2, p_T^2(\pi^0)\right)$ for the $p_T(\pi^0)$ distribution, $\mu^2 = m^2(\pi^0,\pi^0)$ for the $m(\pi^0,\pi^0)$ distribution, and $\mu^2 = p_T^2({\rm jet})$ for the charged hadron distribution. The ``freezing" of the $\pi^0\, p_T$ scale at low $\mu^2$ is introduced in order to prevent PDF grid sampling below its lowest available $Q^2$ value of $(1.65\,{\rm GeV})^2$. 

\begin{figure}[t]
\includegraphics[width=0.49\textwidth]{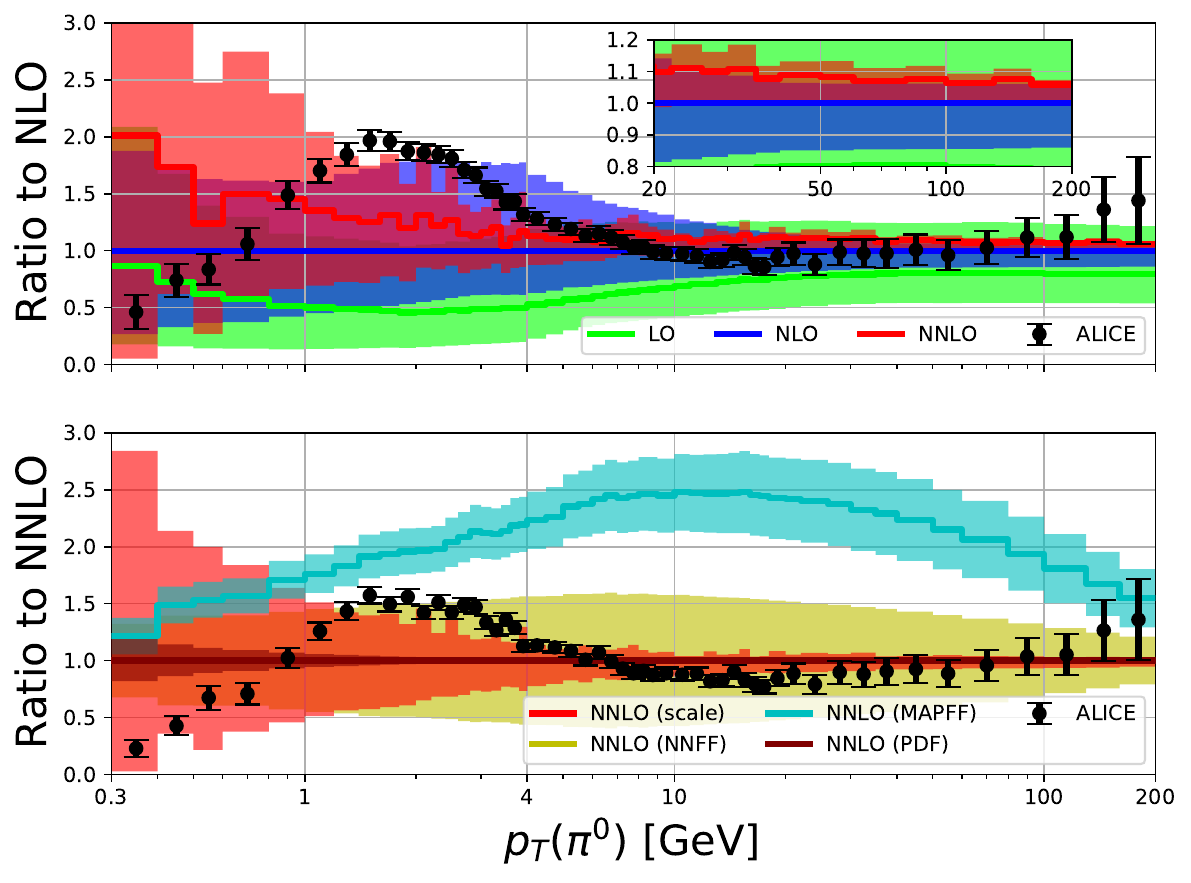}
\caption{Single inclusive $\pi^0$ $p_T$ spectrum at LO (green), NLO (blue)  and NNLO (red) QCD versus ALICE data. Top figure shows ratio to NLO QCD; the colored bands represent scale uncertainty. The inset magnifies the high-$p_T$ region. Bottom figure: two NNLO predictions derived with different FF sets. Shown is a breakdown of PDF, FF and scale uncertainty.}
\label{fig:PT}
\end{figure}

We start by showing in fig.~\ref{fig:PT} the $p_T(\pi^0)$ distribution of a neutral pion in LO, NLO and NNLO QCD. We compare these predictions to $\sqrt{s} = 8$ TeV ALICE data \cite{ALICE:2021est}. The predictions are subject to the selection cut $|y(\pi^0)| < 0.8$. The upper figure shows the theoretical predictions computed at three different orders of perturbation theory, each order showing its own scale dependence. The FF set used in upper figure is NNFF \cite{Bertone:2017tyb}.

We notice that for $p_T$ above 2 GeV, the theoretical predictions exhibit good perturbative convergence in the sense that each subsequent order has a reduced scale variation, the $K$ factors are decreasing and the scale variation bands are overlapping. The NNLO correction increases the NLO prediction but in a manner fully consistent with the NLO uncertainty estimate. The precision of the NNLO prediction increases by a factor of about three, relative to NLO. We also observe that for $p_T$ above about 20 GeV, the scale uncertainty of the NNLO prediction becomes smaller than the overall data uncertainty. This trend becomes even more pronounced for larger $p_T$. 

At lower $p_T$, below about 4 GeV or so, the scale variation of the theory prediction starts to steadily increase. This indicates the onset of the expected breakdown of perturbation theory. Possible improvements, like inclusion of finite-mass effects, have been proposed in the literature \cite{Kneesch:2007ey,Albino:2008fy,MoosaviNejad:2015lgp}. In this work we have not pursued such techniques; instead, we have focused our attention on the behavior of perturbation theory at fixed order.

While the upper plot in fig.~\ref{fig:PT} shows a fairly good theory/data agreement, a meaningful theory/data comparison requires the investigation of all significant sources of theoretical uncertainty. This is done in the lower figure of fig.~\ref{fig:PT} where we have shown the scale and PDF uncertainties of the NNLO prediction, together with the FF uncertainties for two FF sets: NNFF \cite{Bertone:2017tyb} and MAPFF \cite{AbdulKhalek:2022laj}. It is immediately apparent that in the whole $p_T$ range considered in this work the PDF uncertainty is negligible relative to the other sources of theoretical uncertainty. On the other hand, the FF uncertainty dominates over scale uncertainty for all $p_T$ above 2 GeV. In particular, for $p_T(\pi^0)> 6$ GeV, the FF uncertainty exceeds the scale one by a factor of five, or more, making FF variation by far the largest source of theoretical uncertainty. 

The FF uncertainty aside, fig.~\ref{fig:PT} demonstrates that the inclusion of the NNLO correction has important impact, especially in the $p_T$ range between 4 and 30 GeV where the NNLO scale uncertainty is significantly smaller than the NLO one yet not much smaller than the overall data uncertainty.

The significance of the FF uncertainty in this observable can be appreciated in two ways. First, one can compare the FF uncertainty bands of the two different FF sets. As one can see in fig.~\ref{fig:PT} both bands are roughly of the same size. Second, one can compare the difference between the two FF sets. Strikingly, the two sets lead to very different predictions and for $p_T$ above about 1 GeV they appear to be incompatible within their own FF uncertainties. This apparent discrepancy between two modern FF sets indicates that the FF uncertainties of existing FF sets are significant. In order to achieve much improved theory/data comparisons in the future, a new generation of FF sets may be needed. As already mentioned in the Introduction, the calculation presented in this work can be used to derive improved FF fits from hadron collider data.

\begin{figure}[t]
\includegraphics[width=0.49\textwidth]{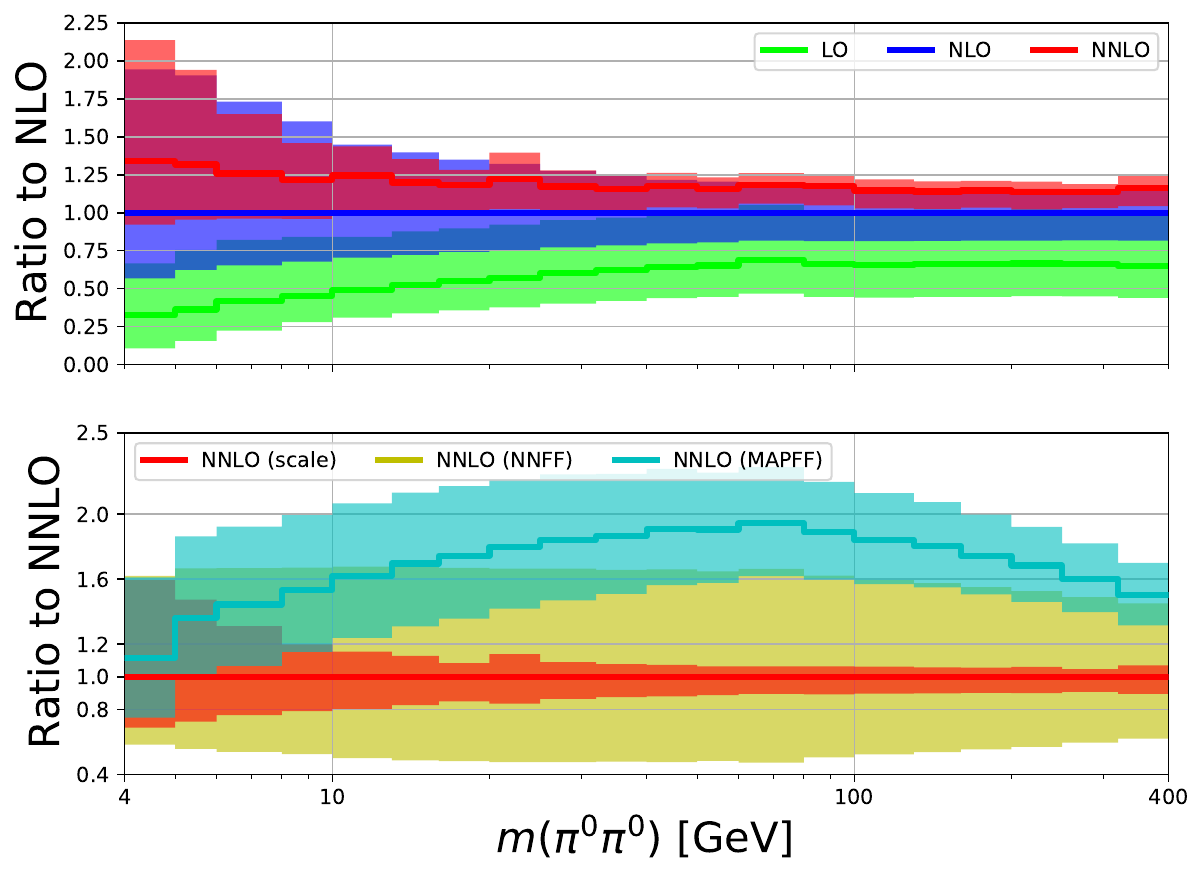}
\caption{As in fig.~\ref{fig:PT} but for the invariant mass of two neutral pions. No comparison to data is shown.}
\label{fig:mpipi}
\end{figure}

In fig.~\ref{fig:mpipi} we present predictions for the $m(\pi^0\pi^0)$ distribution of a pair of neutral pions in LO, NLO and NNLO QCD. No data comparison is available in this case. The prediction is for LHC with $\sqrt{s} = 13$ TeV and is subject to the following selection cuts for each one of the two pions: $|y| < 2.1$ and $p_T > 1$ GeV. 

\begin{figure}[t]
\includegraphics[width=0.45\textwidth]{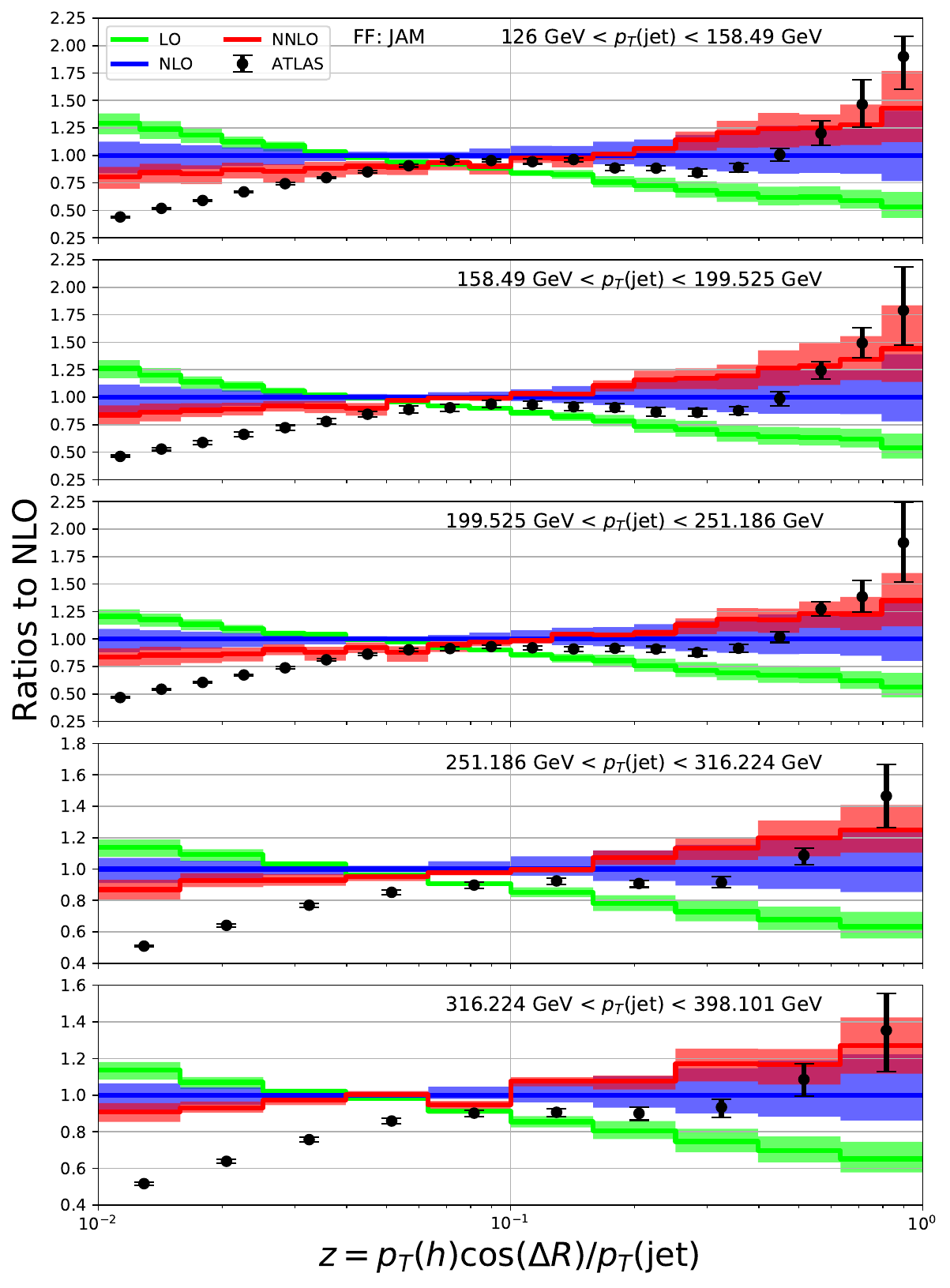}
\caption{The $z$ distribution of charged hadrons inside a jet at LO, NLO and NNLO QCD versus ATLAS data for five separate $p_T({\rm jet})$ slices. Colored bands represent scale uncertainty.}
\label{fig:zjet-orders}
\end{figure}
\begin{figure}[t]
\includegraphics[width=0.45\textwidth]{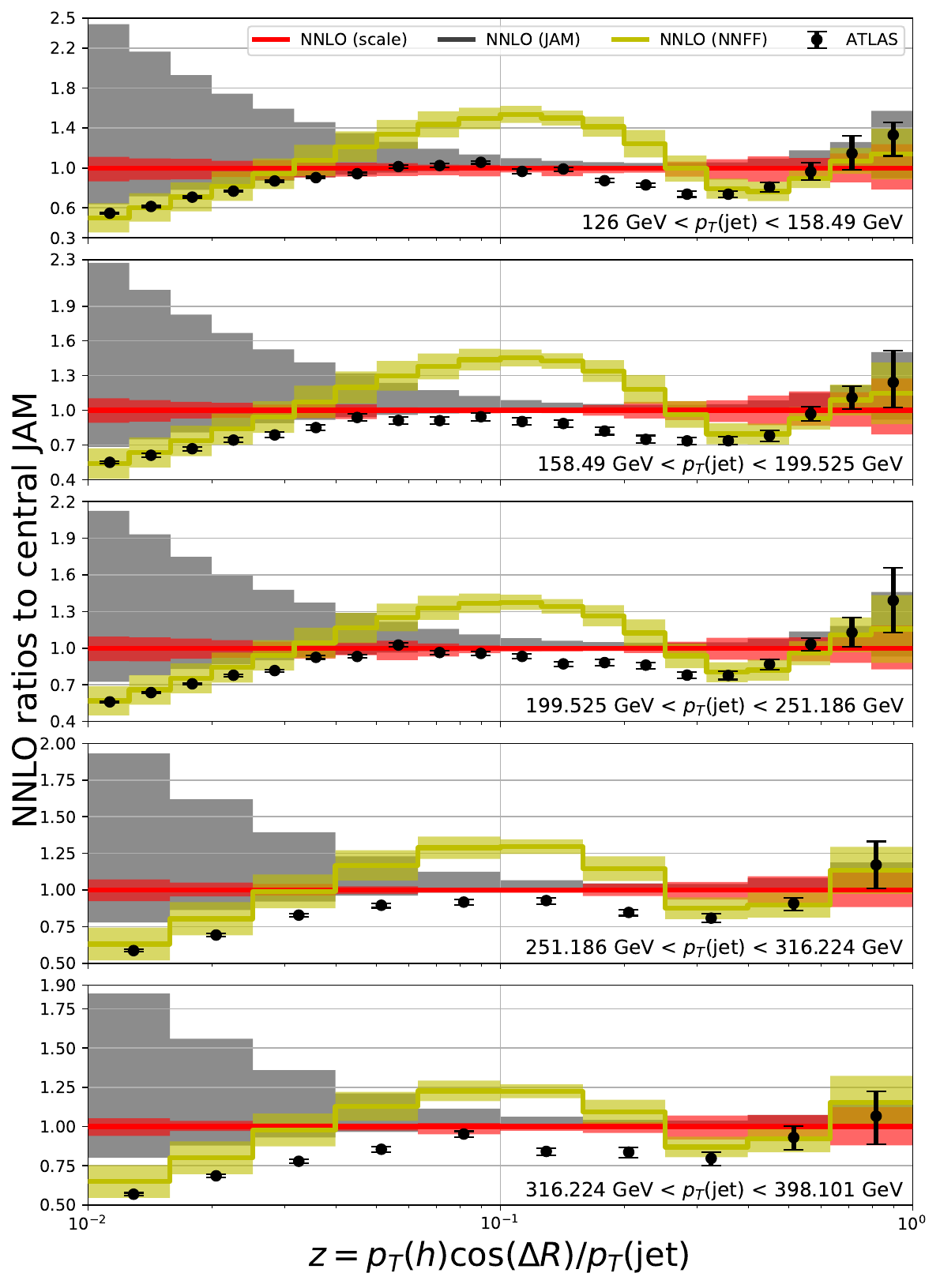}
\caption{As in fig.~\ref{fig:zjet-orders} but at NNLO and for two different FF sets. The bands for each FF set represents its FF uncertainty.}
\label{fig:zjet-FF}
\end{figure}

Overall, the behavior of this observable through NNLO is largely similar to the $p_T(\pi^0)$ distribution discussed above. One observes a consistent $K$-factor and scale dependence reduction  when going from LO to NLO and then to NNLO. The NNLO correction increases the NLO one, while remaining within the NLO scale uncertainty band. The scale variation bands for the three perturbative orders overlap and are consistent with each other. The higher order correction is fairly flat throughout the kinematic range. We conclude that the perturbative convergence for this observable is good and its perturbative description at NNLO is reliable.

As we already noticed in our discussion of the $p_T(\pi^0)$ distribution, the FF uncertainty is by far the largest source of theoretical uncertainty in this observable. The FF uncertainty dominates over the scale one in the full kinematic range, while for $m(\pi^0\pi^0) > 7$ GeV the NNLO scale uncertainty becomes negligible relative to the FF one. A significant difference between the predictions made with the two FF sets, NNFF and MAPFF, can also be observed in this case, although for this observable the two predictions are not inconsistent with each other. This comparison clearly indicates the potential for theoretical improvement in this observable with improved FF sets.

In figs.~\ref{fig:zjet-orders} and \ref{fig:zjet-FF} we show the momentum fraction distribution of a charged hadron inside a jet, $(1/\sigma_{{\rm jet}})d\sigma/dz$ in LO, NLO and NNLO QCD. Jets are clustered with the anti-$k_T$ algorithm \cite{Cacciari:2008gp} with radius $R = 0.4$. We require that $|y({\rm jet})| < 2.1$. We compare our predictions with $\sqrt{s} = 5.02$ TeV ATLAS data \cite{ATLAS:2018bvp}. The data is very precise and extensive, distributed across numerous bins in five $p_T({\rm jet})$ slices. 

In fig.~\ref{fig:zjet-orders} we focus our attention on the theoretical prediction's perturbative convergence through NNLO. The predictions shown in this figure utilize the JAM FF set. The main effect of the inclusion of higher order corrections to this observable is the change in slope of the observed distribution. The scale uncertainty at NLO is larger than at LO, because the LO scale uncertainty is artificially small. This is typical of normalized distributions, where the scale uncertainty mostly drops out in the ratio at LO. At NNLO the scale variation already slightly decreases relative to NLO which indicates the onset of perturbative convergence. The NNLO correction is significant, for all $p_T({\rm jet})$ slices, while remaining consistent with the NLO scale estimate. The overall behavior of this observable is as expected, and it resembles the related distribution for $b$-flavored hadrons studied in great detail in ref.~\cite{Czakon:2021ohs}.

The inclusion of the NNLO correction improves the theory/data comparison for this observable. Nevertheless, it is clear from fig.~\ref{fig:zjet-orders} that the overall theory/data comparison is not satisfactory. Given the significance of the FF uncertainty for the other two observables studied above, in fig.~\ref{fig:zjet-FF} we compare data with NNLO predictions based on two different FF sets: JAM and NNFF. 

The two FF sets lead to strikingly different predictions. For intermediate $z$ values, $0.06 < z < 0.2$, the two are incompatible within their respective FF uncertainties. At large $z$ value the two are compatible, albeit with very different slopes. At low values of $z$ the two are largely consistent although, again, they exhibit very different slopes. As far as data comparison is concerned, neither prediction can satisfactory describe it. Apart form the intermediate $z$ range, the JAM set is consistent with data, while the NNFF set describe data quite well for low and large $z$ but is very far form data in the intermediate $z$ region. The behavior for different $p_T({\rm jet})$ cuts is fairly similar. Noting the very small scale uncertainty of this observable at NNLO, we conclude that with improved future FF sets NNLO QCD has the potential to describe charged hadron data at hadron colliders very precisely.

All predictions presented above are available online from ref.~\cite{website}.

\section{Summary and Outlook}

In this work we have presented the first-ever NNLO QCD calculation of single identified hadrons at hadron colliders. We have observed well-behaved perturbative series at that order with reliable predictions and relatively small residual uncertainty due to missing yet-higher perturbative orders. The largest remaining source of theoretical uncertainty stems from the non-perturbative parton-to-hadron fragmentation functions. The present calculation, especially if combined with the available NNLO calculations for SIA and SIDIS offers the possibility for extracting global FF sets with full NNLO precision.

\begin{acknowledgments}
The work of M.C. was supported by the Deutsche Forschungsgemeinschaft (DFG) under grant 396021762 - TRR 257: Particle Physics Phenomenology after the Higgs Discovery. T.G. and A.M. have been supported by STFC consolidated HEP theory grants ST/T000694/1 and ST/X000664/1. This work was performed using the Cambridge Service for Data Driven Discovery (CSD3), part of which is operated by the University of Cambridge Research Computing on behalf of the STFC DiRAC HPC Facility (www.dirac.ac.uk). The DiRAC component of CSD3 was supported by STFC grants ST/P002307/1, ST/R002452/1 and ST/R00689X/1.
\end{acknowledgments}


\begin{thebibliography}{99}

\bibitem{Aversa:1988vb}
F.~Aversa, P.~Chiappetta, M.~Greco and J.~P.~Guillet,
Nucl. Phys. B \textbf{327}, 105 (1989)

\bibitem{Jager:2002xm}
B.~Jager, A.~Schafer, M.~Stratmann and W.~Vogelsang,
Phys. Rev. D \textbf{67}, 054005 (2003)
[arXiv:hep-ph/0211007 [hep-ph]].

\bibitem{Hinderer:2018nkb}
P.~Hinderer, F.~Ringer, G.~Sterman and W.~Vogelsang,
``Threshold Resummation at NNLL for Single-particle Production in Hadronic Collisions,''
Phys. Rev. D \textbf{99}, no.5, 054019 (2019)
[arXiv:1812.00915 [hep-ph]].

\bibitem{Arleo:2013tya}
F.~Arleo, M.~Fontannaz, J.~P.~Guillet and C.~L.~Nguyen,
JHEP \textbf{04}, 147 (2014)
[arXiv:1311.7356 [hep-ph]].

\bibitem{Kaufmann:2019ksh}
T.~Kaufmann, X.~Liu, A.~Mukherjee, F.~Ringer and W.~Vogelsang,
``Hadron-in-jet production at partonic threshold,''
JHEP \textbf{02}, 040 (2020)
[arXiv:1910.11746 [hep-ph]].

\bibitem{Liu:2023fsq}
C.~Liu, X.~Shen, B.~Zhou and J.~Gao,
JHEP \textbf{09}, 108 (2023)
[arXiv:2305.14620 [hep-ph]].

\bibitem{Caletti:2024xaw}
S.~Caletti, A.~Gehrmann-De Ridder, A.~Huss, A.~R.~Garcia and G.~Stagnitto,
JHEP \textbf{10}, 027 (2024)
[arXiv:2405.17540 [hep-ph]].

\bibitem{Hirai:2007cx}
M.~Hirai, S.~Kumano, T.~H.~Nagai and K.~Sudoh,
Phys. Rev. D \textbf{75}, 094009 (2007)
[arXiv:hep-ph/0702250 [hep-ph]].

\bibitem{deFlorian:2007ekg}
D.~de Florian, R.~Sassot and M.~Stratmann,
Phys. Rev. D \textbf{76}, 074033 (2007)
[arXiv:0707.1506 [hep-ph]].

\bibitem{Albino:2008fy}
S.~Albino, B.~A.~Kniehl and G.~Kramer,
Nucl. Phys. B \textbf{803}, 42-104 (2008)
[arXiv:0803.2768 [hep-ph]].

\bibitem{deFlorian:2014xna}
D.~de Florian, R.~Sassot, M.~Epele, R.~J.~Hern\'andez-Pinto and M.~Stratmann,
Phys. Rev. D \textbf{91}, no.1, 014035 (2015)
[arXiv:1410.6027 [hep-ph]].

\bibitem{Bertone:2017tyb}
V.~Bertone \textit{et al.} [NNPDF],
Eur. Phys. J. C \textbf{77}, no.8, 516 (2017)
[arXiv:1706.07049 [hep-ph]].

\bibitem{Bertone:2018ecm}
V.~Bertone \textit{et al.} [NNPDF],
Eur. Phys. J. C \textbf{78}, no.8, 651 (2018)
[erratum: Eur. Phys. J. C \textbf{84}, no.2, 155 (2024)]
[arXiv:1807.03310 [hep-ph]].

\bibitem{Moffat:2021dji}
E.~Moffat \textit{et al.} [Jefferson Lab Angular Momentum (JAM)],
Phys. Rev. D \textbf{104}, no.1, 016015 (2021)
[arXiv:2101.04664 [hep-ph]].

\bibitem{Khalek:2021gxf}
R.~A.~Khalek \textit{et al.} [MAP (Multi-dimensional Analyses of Partonic distributions)],
Phys. Rev. D \textbf{104}, no.3, 034007 (2021)
[arXiv:2105.08725 [hep-ph]].

\bibitem{Borsa:2021ran}
I.~Borsa, D.~de Florian, R.~Sassot and M.~Stratmann,
Phys. Rev. D \textbf{105}, no.3, L031502 (2022)
[arXiv:2110.14015 [hep-ph]].

\bibitem{Borsa:2022vvp}
I.~Borsa, R.~Sassot, D.~de Florian, M.~Stratmann and W.~Vogelsang,
Phys. Rev. Lett. \textbf{129}, no.1, 012002 (2022)
[arXiv:2202.05060 [hep-ph]].

\bibitem{AbdulKhalek:2022laj}
R.~Abdul Khalek \textit{et al.} [MAP (Multi-dimensional Analyses of Partonic distributions)],
Phys. Lett. B \textbf{834}, 137456 (2022)
[arXiv:2204.10331 [hep-ph]].

\bibitem{Gao:2024dbv}
J.~Gao, C.~Liu, X.~Shen, H.~Xing and Y.~Zhao,
Phys. Rev. D \textbf{110}, no.11, 114019 (2024)
[arXiv:2407.04422 [hep-ph]].

\bibitem{Goyal:2023zdi}
S.~Goyal, S.~O.~Moch, V.~Pathak, N.~Rana and V.~Ravindran,
Phys. Rev. Lett. \textbf{132}, no.25, 251902 (2024)
doi:10.1103/PhysRevLett.132.251902
[arXiv:2312.17711 [hep-ph]].

\bibitem{Bonino:2024qbh}
L.~Bonino, T.~Gehrmann and G.~Stagnitto,
Phys. Rev. Lett. \textbf{132}, no.25, 251901 (2024)
[arXiv:2401.16281 [hep-ph]].

\bibitem{Bonino:2024wgg}
L.~Bonino, T.~Gehrmann, M.~L\"ochner, K.~Sch\"onwald and G.~Stagnitto,
Phys. Rev. Lett. \textbf{133}, no.21, 211904 (2024)
[arXiv:2404.08597 [hep-ph]].

\bibitem{Goyal:2024tmo}
S.~Goyal, R.~N.~Lee, S.~O.~Moch, V.~Pathak, N.~Rana and V.~Ravindran,
Phys. Rev. Lett. \textbf{133}, 211905 (2024)
doi:10.1103/PhysRevLett.133.211905
[arXiv:2404.09959 [hep-ph]].

\bibitem{Ahmed:2024owh}
T.~Ahmed, S.~Goyal, S.~M.~Hasan, R.~N.~Lee, S.~O.~Moch, V.~Pathak, N.~Rana, A.~Rapakoulias and V.~Ravindran,
[arXiv:2412.16509 [hep-ph]].

\bibitem{Goyal:2024emo}
S.~Goyal, R.~N.~Lee, S.~O.~Moch, V.~Pathak, N.~Rana and V.~Ravindran,
[arXiv:2412.19309 [hep-ph]].

\bibitem{Buckley:2014ana}
A.~Buckley, J.~Ferrando, S.~Lloyd, K.~Nordstr\"om, B.~Page, M.~R\"ufenacht, M.~Sch\"onherr and G.~Watt,
Eur. Phys. J. C \textbf{75}, 132 (2015)
[arXiv:1412.7420 [hep-ph]].

\bibitem{NNPDF:2017mvq}
R.~D.~Ball \textit{et al.} [NNPDF],
Eur. Phys. J. C \textbf{77}, no.10, 663 (2017)
[arXiv:1706.00428 [hep-ph]].

\bibitem{Czakon:2024tjr}
M.~Czakon, T.~Generet, A.~Mitov and R.~Poncelet,
[arXiv:2411.09684 [hep-ph]].

\bibitem{Czakon:2010td}
M.~Czakon,
Phys. Lett. B \textbf{693}, 259-268 (2010)
[arXiv:1005.0274 [hep-ph]].

\bibitem{Czakon:2011ve}
M.~Czakon,
Nucl. Phys. B \textbf{849}, 250-295 (2011)
[arXiv:1101.0642 [hep-ph]].

\bibitem{Czakon:2014oma}
M.~Czakon and D.~Heymes,
Nucl. Phys. B \textbf{890}, 152-227 (2014)
[arXiv:1408.2500 [hep-ph]].

\bibitem{Czakon:2019tmo}
M.~Czakon, A.~van Hameren, A.~Mitov and R.~Poncelet,
JHEP \textbf{10}, 262 (2019)
[arXiv:1907.12911 [hep-ph]].

\bibitem{Czakon:2021ohs}
M.~Czakon, T.~Generet, A.~Mitov and R.~Poncelet,
JHEP \textbf{10}, 216 (2021)
[arXiv:2102.08267 [hep-ph]].

\bibitem{Broggio:2014hoa}
A.~Broggio, A.~Ferroglia, B.~D.~Pecjak and Z.~Zhang,
JHEP \textbf{12}, 005 (2014)
[arXiv:1409.5294 [hep-ph]].

\bibitem{Bern:2002tk}
Z.~Bern, A.~De Freitas and L.~J.~Dixon,
JHEP \textbf{03}, 018 (2002)
[arXiv:hep-ph/0201161 [hep-ph]].

\bibitem{Bern:2003ck}
Z.~Bern, A.~De Freitas and L.~J.~Dixon,
JHEP \textbf{06}, 028 (2003)
[erratum: JHEP \textbf{04}, 112 (2014)]
[arXiv:hep-ph/0304168 [hep-ph]].

\bibitem{Glover:2003cm}
E.~W.~N.~Glover and M.~E.~Tejeda-Yeomans,
JHEP \textbf{06}, 033 (2003)
[arXiv:hep-ph/0304169 [hep-ph]].

\bibitem{Glover:2004si}
E.~W.~N.~Glover,
JHEP \textbf{04}, 021 (2004)
[arXiv:hep-ph/0401119 [hep-ph]].

\bibitem{DeFreitas:2004kmi}
A.~De Freitas and Z.~Bern,
JHEP \textbf{09}, 039 (2004)
[arXiv:hep-ph/0409007 [hep-ph]].

\bibitem{Buccioni:2019sur}
F.~Buccioni, J.~N.~Lang, J.~M.~Lindert, P.~Maierh\"ofer, S.~Pozzorini, H.~Zhang and M.~F.~Zoller,
Eur. Phys. J. C \textbf{79}, no.10, 866 (2019)
[arXiv:1907.13071 [hep-ph]].

\bibitem{Bury:2015dla}
M.~Bury and A.~van Hameren,
Comput. Phys. Commun. \textbf{196}, 592-598 (2015)
[arXiv:1503.08612 [hep-ph]].

\bibitem{Gao:2025bko}
J.~Gao, C.~Liu, M.~Li, X.~Shen, H.~Xing, Y.~Zhao and Y.~Zhou,
[arXiv:2503.21311 [hep-ph]].

\bibitem{ALICE:2021est}
S.~Acharya \textit{et al.} [ALICE],
Phys. Lett. B \textbf{827}, 136943 (2022)
[arXiv:2104.03116 [nucl-ex]].

\bibitem{Kneesch:2007ey}
T.~Kneesch, B.~A.~Kniehl, G.~Kramer and I.~Schienbein,
Nucl. Phys. B \textbf{799}, 34-59 (2008)
[arXiv:0712.0481 [hep-ph]].

\bibitem{MoosaviNejad:2015lgp}
S.~M.~Moosavi Nejad, M.~Soleymaninia and A.~Maktoubian,
Eur. Phys. J. A \textbf{52}, no.10, 316 (2016)
[arXiv:1512.01855 [hep-ph]].

\bibitem{Cacciari:2008gp}
M.~Cacciari, G.~P.~Salam and G.~Soyez,
JHEP \textbf{04}, 063 (2008)
[arXiv:0802.1189 [hep-ph]].

\bibitem{ATLAS:2018bvp}
M.~Aaboud \textit{et al.} [ATLAS],
Phys. Rev. C \textbf{98}, no.2, 024908 (2018)
[arXiv:1805.05424 [nucl-ex]].

\bibitem{website}
\href{https://www.precision.hep.phy.cam.ac.uk/results/fragmentation-functions/}{https://www.precision.hep.phy.cam.ac.uk/results/\\fragmentation-functions/}



\end{thebibliography}
\end{document}